\documentclass[journal=acs-an,manuscript=article]{achemso}

\usepackage[T1]{fontenc}

\usepackage{xcolor}
\definecolor{col_black}{RGB}{0,0,0}
\definecolor{col_review1}{RGB}{0,0,0} 
\definecolor{col_review2}{RGB}{0,0,0} 
\newcommand{\reviewfirst}[1]{\textcolor{col_review1}{#1}}
\newcommand{\reviewsecond}[1]{\textcolor{col_review2}{#1}}

\usepackage{breakurl}
\usepackage{natbib}
\usepackage{url}

\author{Maxime Vassaux}
\email{maxime.vassaux@univ-rennes.fr}
\affiliation[ipr]{Univ. Rennes, CNRS, IPR - UMR 6251, Rennes, 35000, France}

\title{Heterogeneous structure and dynamics of water in a hydrated collagen microfibril}

\begin{document}

\maketitle

\begin{abstract}
\reviewfirst{Collagen type I is well-known for its outstanding mechanical properties which it inherits from its hierarchical structure. Collagen type I fibrils may be viewed as an \reviewsecond{heterogeneous material} made of protein, macromolecules (such as glycosaminoglycans and proteoglycans) and water. Water content modulates the properties of these fibrils.} Yet, the properties of water and the fine interactions of water with the protein constituent of these \reviewsecond{heterofibrils} have only received limited attention. \reviewfirst{Here, we propose to model collagen type I fibrils as an hydrated structure made of tropocollagen molecules assembled in a microfibril crystal. We perform large-scale all-atom molecular dynamics simulations of the hydration of collagen fibrils beyond the onset of disassembly.} We found that the structural and dynamic properties of water vary strongly with the level of hydration of the microfibril. More importantly, we found that the properties vary spatially within the $67$ nm D-spacing periodic structure. \reviewfirst{\reviewsecond{Alteration} of the structural and dynamical properties of the collagen microfibril occur first in the gap region. Overall, we identify that the change in the role of water molecules from glue to lubricant between tropocollagen molecules arises around $100$\% hydration while the microfibril begins to disassemble beyond $130$\% water content. Our findings are supported by a decrease in hydrogen bonding, recovery of bulk water properties and amorphisation of the tropocollagen molecules packing. Our simulations reveal the structure and dynamics of hydrated collagen fibrils with unprecedented spatial resolution from physiological conditions to disassembly. Beyond the process of self-assembly and the emergence of mechanical properties of collagen type I fibrils, our results may also provide new insights in mineralization of collagen fibrils.}
\end{abstract}


\section{Keywords}
collagen type I, nanoconfined water, hydration, molecular dynamics, fibrils self-assembly

\section{Introduction}
\label{sec:intro}
\reviewfirst{Collagen is ubiquitous in the animal world, from fishes to mammals, it is the protein of choice of organisms to design structural materials. Collagen type I is the most occurring type of collagen among the 28 co-existing collagen sequences, it is found in tendons, bone and cornea among others \cite{ricard-blum_collagen_2011}. Collagen-based tissues feature a wide-range of outstanding mechanical properties (elastic, viscous, fracture) which are tuned modifying the hierarchical fibrillar structure of collagen self-assemblies \cite{andriotis_mechanics_2023}. Tuning of properties often relies on the association of collagen with additional organic (e.g. in tendon, ligaments) or inorganic compounds (e.g. in bone) \cite{fratzl_collagen_2008}. The specific, amino acid structure, dictating the self-assembly of tropocollagen molecules into this intricate fibrillar architecture is widely accepted as a key ingredient of the mechanical properties of the tissues made of collagen type I \cite{fratzl_fibrillar_1998}. One significantly abundant collagen self-assembled structure is the collagen fibril which consists of twisted bundles of aligned tropocollagen molecules. The fibrillar structure of collagen itself is mostly correlated with its characteristic repetitive sequence of amino acids. The presence of additional molecules is key to the stability of these fibrillar self-assemblies \cite{hulmes_collagen_2008}, macromolecules in particular, such as proteoglycans or glycosaminoglycans, cross-links and last but not least water molecules.}\\

The structure of water molecules within the the collagen fibril, around and in between tropocollagen molecules, has been characterised using nuclear magnetic resonance (NMR) \cite{fullerton_evidence_2006} \reviewfirst{and very recently using atomic force microscopy (AFM) \cite{arvelo_interfacial_2024}. The dynamics of water molecules within the fibril have also been accessed using NMR showing two orders of magnitude slower diffusion of collagen water than bulk water \cite{masiewicz_dynamical_2024}. Hydrogen bonding mediated by water molecules at varying thermodynamics conditions has been investigated using molecular dynamics, highlighting the stability of hydrogen bond network irrespective of temperature \cite{madhavi_structure_2019}.}\\

Hydration is known to be key for collagen stability \cite{mogilner_collagen_2002}. \reviewfirst{The mechanical properties of collagen fibrils such as Young's modulus can drop down by two orders of magnitude depending on their water content \cite{harley_phonons_1977, wenger_mechanical_2007, grant_effects_2008}, and upon drying fibrils become extensively brittle \cite{shen_vitro_2010}. Water molecules within the fibril mediate forces between tropocollagen molecules, intermolecular distances have been closely related to water content \cite{leikin_direct_1994, leikin_raman_1997, gautieri_hydration_2012}. Removal of water molecules from the fibril structure has been shown to lead to heterogeneous shrinking as well as the emerge of severe tensile stresses \cite{masic_osmotic_2015}.}\\

\reviewfirst{When it comes to finely varying hydration, and therefore the mass ratio of water to protein, data remain scarce, limited to conformational changes of tropocollagen molecules. The difficulty of controlling water content within collagen fibrils experimentally have limited access to quantitative analysis of the influence of hydration on water properties. Varying solvent chemistry, increasing osmotic pressure on collagen fibrils has been used in most studies of collagen at varying water content \cite{leikin_direct_1994}. Experimental studies were therefore limited on water contents mostly ranging from dry to $80$\% water-to-protein mass ratio, which correspond to a full range of relative humidity ($0$\% to $100$\%). In addition to intermolecular forces, swelling of the fibrils was characterised using AFM \cite{grant_tuning_2009, spitzner_nanoscale_2015, andriotis_hydration_2019}. Heavy water (D$_2$O) has also been used a way to fine tune hydration of fibrils modulating hydrogen bonding, which resulted in lower hydration and faster self-assembly \cite{giubertoni_elucidating_2024}. Temperature has also been used to trigger structural changes in collagen fibrils and influence intermolecular forces \cite{leikin_temperature-favoured_1995}.}\\ 

\reviewfirst{Beyond structural aspects, water dynamics within the fibril have received no or at most very little attention. Yet, both largely affect conformation and functionalisation of proteins \cite{laage_water_2017}.} To that extent, an analysis of the effect of hydration, that is water structure and dynamics, on the molecular structural and mechanical properties of collagen fibrils is still lacking. \reviewfirst{In particular, with spatial resolution within the fibril structure and over a wide range of hydration contents.} Such understanding is key to understand the emergence of physical properties of collagen fibrils and collagen-based tissues.\\

\reviewfirst{We here propose a thorough investigation of the structure and dynamics of water molecules confined within the collagen microfibril crystal. The microfibril crystal is a repetitive subunit found in collagen fibrils made of fragments of five tropocollagen molecules \cite{orgel_microfibrillar_2006}. Within the microfibril can be found the overlap and gap regions constituting the periodic D-spacing of collagen fibrils. We here consider a microfibril crystal constituted of tropocollagen and water molecules, excluding all other macromolecules}. \reviewfirst{Our investigation relies on long all-atom molecular dynamics simulation of the hydration of tropocollagen molecules, from the water-to-protein mass ratio of tendon (approx. $60$\%)\cite{fullerton_orientation_1985} to large hydration degree, beyond the supposed onset of the fibril disassembly (approx. $250$\%).} And we focus on providing spatially-resolved details about water structure (radial distribution function, hydrogen bonds) and dynamics (self-diffusion coefficient), and protein structure throughout the microfibril crystal, and in particular in the characteristic overlap and gap regions (see Figure \ref{fig:molecular_model}.a).

\section{Methods}
\label{sec:methods}

Molecular dynamics simulations of the hydrated collagen microfibril are performed using the massively parallel NAMD package \cite{phillips_scalable_2020}. \reviewfirst{The molecular model only considers atom for the tropocollagen molecules, couterions and water molecules}. The potentials describing amino acid atoms interactions are modeled and parameterised with the Generic Amber Force Field (GAFF) \cite{wang_development_2004} which has been extensively used for the simulation of protein structure and functionalisation. Meanwhile the potentials describing the interactions involving water atoms rely on the TIP3P force field \cite{jorgensen_comparison_1983}, we use a rigid water model for computational efficiency, as we do not intend to focus on high-frequency bond and angle vibrational modes. We perform ensemble MD simulations for better precision of the thermodynamic averages, the ensemble features $6$ replicas with different randomly initialised velocity of the protein atoms.\\

The position of the protein backbone atoms is set according to the microfibril crystallographic structure obtained by \citep{orgel_microfibrillar_2006} using X-ray fiber diffraction on native rat tail tendon \textit{in situ}. The structure labelled \textit{3HR2} containing the position of the atoms can be found on the Protein Data Bank \cite{berman_protein_2000}. The molecular system is set up using periodic boundary conditions and the microfibril crystals triclinic box \cite{orgel_microfibrillar_2006}. We modified amino acids from the \textit{3HR2} structure to match the exact sequence of human collagen type I as found in the single chains' sequences denoted \textit{NP\_000079} (for alpha-1(I) helices) and \textit{NP\_000080} (for the alpha-2(I) helices). Nonetheless, we preserved hydroxyproline amino acids in the sequence of the chains according to the original structure. Last, we added $34$ negatively charged chlorine ions to neutralise the system, which interactions are parameterised using GAFF.\\

\begin{figure}
    \centering
    \includegraphics[width=\textwidth]{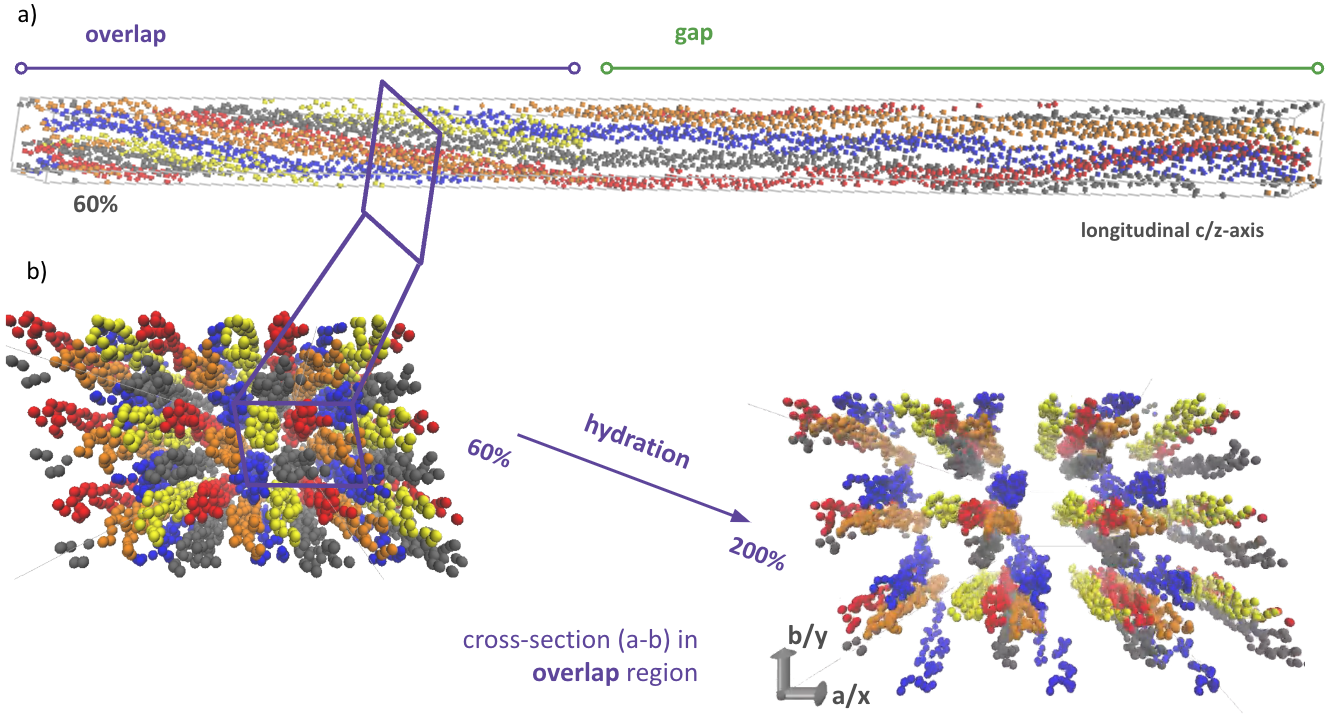}
    \caption{\textbf{Molecular model of the hydrated microfibril.} (a) Visualisation of the complete microfibril structure at $60$\% hydration illustrating the overlap (5-stranded) and gap (4-stranded) regions.(b) Visualisation of a $10$ nm-thick cross section of the microfibril in the overlap region at $60$\% and $200$\% hydration. Note that in the cross section, 3x3 unit cells (light blue) are shown. Water molecules have been removed for clarity. Each strand of tropocollagen is colored specifically.}
    \label{fig:molecular_model}
\end{figure}

Once the initial structure of the protein is set, we proceed to hydration of the molecular system in two stages following procedures found in the literature \cite{streeter_molecular_2011,gautieri_hierarchical_2011}. First, we add water molecules such as to stabilise the volume of the microfibril crystal characterised by X-ray crystallography. \reviewfirst{This is assumed to be the amount of water present in the tendon at physiological conditions \cite{fullerton_orientation_1985}. Second, we iteratively add a fixed amount of water molecules up to a hydration level equal to approximately $250$\% water-to-protein mass ratio at which we may expect disassembly.} Water molecules are inserted at random location in the microfibril crystal such that all water molecules are distant of the protein atoms from at least $0.3$ nm and of the other water molecules from at least $0.1$ nm. This is the standard \textit{addWater} procedure found in the AmberTools suite \cite{case_amber_2005}. After each addition of water molecules, whether it is to stabilise or to progressively hydrate the crystal, we perform the following equilibration simulation in three stages: (i) a $10000$ iterations molecular mechanics minimisation of the potential energy with fixed protein backbone atoms; (ii) a $0.5$ ns molecular dynamics simulation at fixed volume and at temperature set at 298 K (\textit{NVT} ensemble) using the Langevin thermostat (damping coefficient: $100$ fs), and with fixed protein backbone atoms; and (iii) a $10$ ns molecular dynamics simulation at pressure set at $0.1$ MPa and at temperature set at $298$ K (\textit{NPT} ensemble) using respectively the isotropic Berendsen barostat (pressure relaxation time: $100$ fs) and the Langevin thermostat (damping coefficient: $100$ fs).\\

Note that for both molecular dynamics simulations, assuming that the molecular system is at equilibrium, we draw the initial velocities of the atoms randomly from the Maxwell–Boltzmann distribution parameterised at $298$ K.\\

The first stage of the hydration process is achieved in a trial-and-error fashion, we insert a given number of water molecules in the microfibril crystal and we check whether the volume of the crystal remains stable at $726$ nm$^3$, the volume obtained experimentally \cite{orgel_microfibrillar_2006}. We found that $12550$ water molecules is the water content which stabilises the microfibril crystal. Such water content is approximately equivalent to \citet{streeter_molecular_2011} and $10$\% higher than to \citet{gautieri_hierarchical_2011}. One of the main difference with the later molecular model is the use of the force field, \citet{gautieri_hierarchical_2011} used the GROMOS 43a1 force field from the GROMACS 4.0 code \cite{berendsen_gromacs_1995}.\\

After the first stage, our equilibrated molecular system features a water-to-protein mass ratio about $57$\%. The subsequent second stage of the hydration process is achieved iteratively adding a fixed amount of water molecules up to $250$\% ($56550$ water molecules). The iterative procedure follows three stages: (i) an insertion of $1000$ water molecules; (ii) the aforementioned equilibration simulation workflow; and (iii) an additional molecular dynamics simulation at pressure set at $0.1$ MPa and at temperature set at $298$ K (\textit{NPT} ensemble) using respectively the anisotropic Berendsen barostat (pressure relaxation time: $100$ fs) and the Langevin thermostat (damping coefficient: $100$ fs), until the long dimension of the microfibril relaxes to its characteristic experimental length $67$ nm.\\

These steps are repeated until reaching the desired hydration level. Note that the simulation in stage (iii) using an anisotropic barostat, allowing non-proportional deformation of the microfibril crystal, is necessary to recover the observed characteristic length of the microfibril crystal, the so-called \textit{D-spacing}, approximately $67$ nm \cite{hodge_effects_1960,tromans_electron_1962,olsen_electron_1963}. This stage systematically lasts less than $1$ ps. The molecular model of the hydrated microfibril is shown in Figure \ref{fig:molecular_model}. \reviewfirst{The application of such a constraint on the $z$-axis length of the microfibril crystal alleviate the lack of stability of tropocollagen molecules \cite{leikina_type_2002}. The constraint is not expected to modify the \reviewsecond{swelling} response of collagen to hydration. \reviewsecond{Hydration has only been shown to influence the gap-to-overlap regions ratio, whereas the influence on the \textit{D-spacing} remained small (about $2$\%) and radial changes were insensitive to longitudinal variations constraints\cite{masic_osmotic_2015}.}}\\

The verification of the equilibration of the initial molecular system (energy and volume, see figure S1.a), as well as the evolution of the dimensions of the microfibril crystal at different hydration levels (see figure S1.b) are thoroughly illustrated in the supplementary information.\\

Analysis of the water and protein data (structure and dynamics) has been facilitated by the use of several modules from the Python library \textit{MDAnalysis} \cite{michaud_mdanalysis_2011,gowers_mdanalysis_2016}.\\ 
\reviewfirst{Computations of radial distribution functions (RDF or $g(r)$) between two groups of atoms $a$ and $b$ which counts the average number of b neighbours in a shell at distance r around a particle $a$ using the following formula:
\begin{equation}
    g_{ab}(r) = \frac{1}{\rho N_{a} N_{b}} \sum_{i=1}^{N_a} \sum_{j=1}^{N_b} \delta(|\mathbf{r}_i - \mathbf{r}_j| - r)
\end{equation}
where $N_{a,b}$ is the number of atoms in a given group, $\rho$ the density of the system, $\mathbf{r}_{i}$ is the vector position of atom $i$ and the function $\delta$ which is equal to $1$ if its argument is $0$, and $0$ elsewhere.\\}
\reviewfirst{Computations of time-averaged mean square displacements (MSD) using the following formula:
\begin{equation}
MSD(r) = \bigg{\langle} \frac{1}{N} \sum_{i=1}^{N} |r(t) - r(t_0)|^2 \bigg{\rangle}_{t_{0}}  
\end{equation}
where $N$ is the number of water molecules, $r$ is their coordinates and $t$ the lag time from $t_0$, $t_0$ being varied for better sampling.\\}
\reviewfirst{Last, computations of hydrogen bonds rely the following criteria: (i) a cutoff maximum distance of $1.2$ \AA~between the donor and the hydrogen atoms, (ii) a cutoff maximum distance of $3$ \AA~between the donor and the acceptor atoms, and (iii) a cutoff minimum angle of $150$° between the donor, hydrogen and acceptor atoms. Hydrogen bonds are classified based on the type of acceptor and donor molecules, whether it is a water molecule (\textit{w}) or a protein (\textit{p}).}

\section{Results}

\begin{figure}
    \centering
    \includegraphics[width=\textwidth]{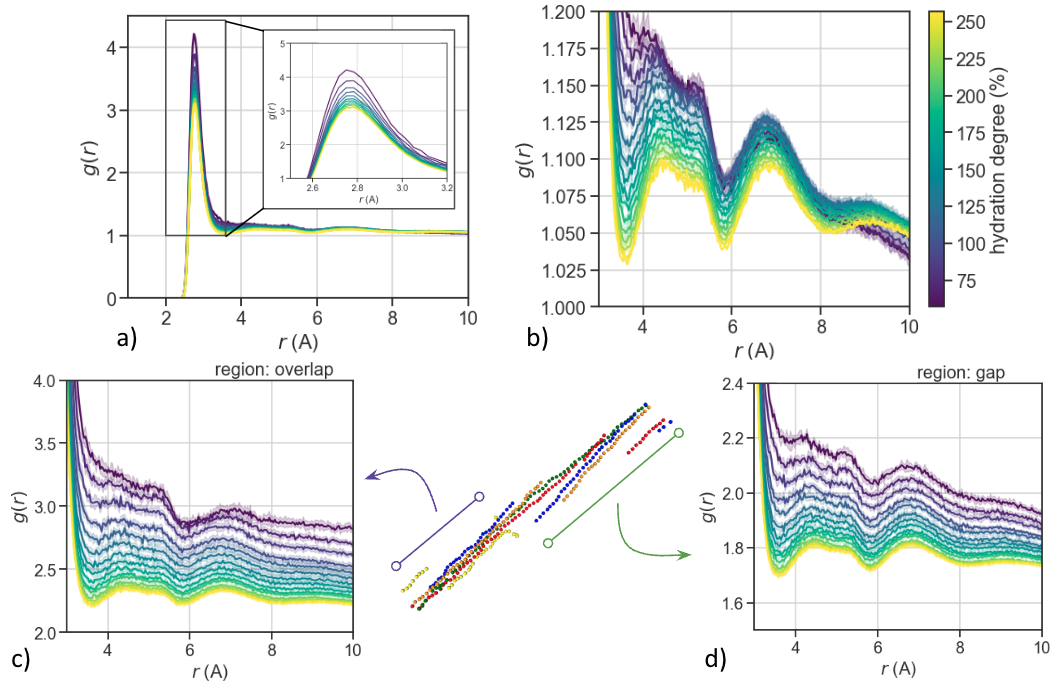}
    \caption{\textbf{Water structure in the collagen microfibril.} (a) Evolution of the radial distribution function ($g(r)$) of water to water atoms with the hydration degree; (b) zoom on the structure of water at medium and long-range, showing much clearer medium-range order at high level of hydration; further detail provided for (c) the overlap and (d) the gap regions show that the structuration of water is more significant in the gap region. \reviewfirst{Error bars are obtained from time-averaging over the last $2$ ns of the simulation and from ensemble-averaging over $6$ replicas.}}
    \label{fig:water_structure}
\end{figure}

We first focus on the structure of water throughout the microfibril. We compute the RDF $g(r)$ of the oxygen atoms in water molecules (see Figure \ref{fig:water_structure}). \reviewsecond{The relative intensity of peaks within a single RDF is the main quantity of interest, as well as their width and their position. The width or sharpness indicates the degree of ordering of the structure.} The structure of water molecules on average over the whole microfibril crystal, but also over the overlap and over the gap regions becomes more organised for higher hydration levels (see Figure \ref{fig:water_structure}.b,c,d, respectively). A separation of the first and the second peak appears clearer beyond $100$\% hydration. \reviewsecond{The second peak (around $4.5$ nm and ending at $5.8$ nm) of $g(r)$ appears progressively sharper and more clearly separated from the first peak. The shape of third peak ($6.7$ nm and ending at $8.1$ nm) does not change, but its relative intensity increases slightly.} Besides, we observe a decrease of amplitude and shift of the first peak with increasing hydration (from $2.75$ to $2.79$ \AA) (see inset in Figure \ref{fig:water_structure}.a). Such an an increase of the distance between neighbouring water molecules can be associated with a decrease in confinement and the swelling of the microfibril (see Figure \ref{fig:molecular_model}.b). Water molecules are getting organised in a more structured but less dense packing.\\

\begin{figure}
    \centering
    \includegraphics[width=\textwidth]{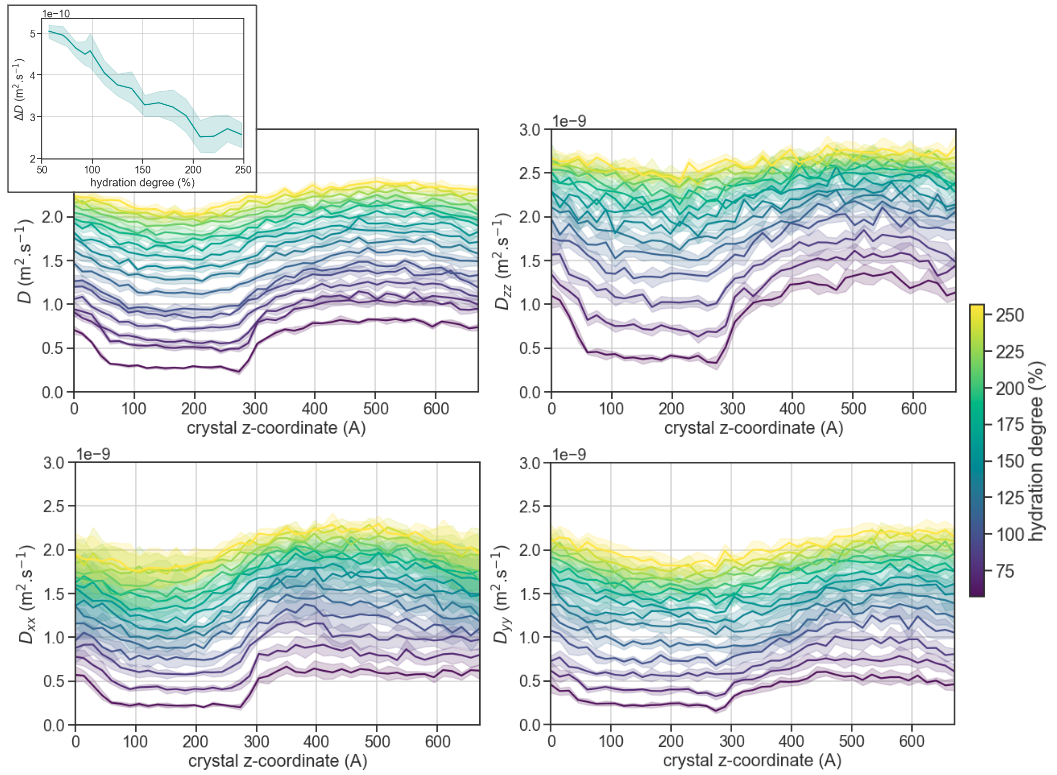}
    \caption{\textbf{Diffusion of water molecules in the collagen microfibril.} Evolution of the total ($D$) and directional ($D_{xx}$, $D_{yy}$ or $D_{zz}$) self-diffusion coefficients for varying positions along the $z$-axis of the microfibril crystal and for varying hydration levels (color scale). \reviewfirst{Error bars are obtained from time-averaging over the last $2$ ns of the simulation and from ensemble-averaging over $6$ replicas.} \reviewfirst{The inset in the top left panel shows the evolution with the hydration degree of the difference of self-diffusion coefficient between the gap and the overlap region ($\Delta D$).}}
    \label{fig:water_dynamics}
\end{figure}

We then focus on the dynamics of water. The self-diffusion coefficient computed from the MSD of individual water molecules within the collagen microfibril is used for analysis (see Figure \ref{fig:water_dynamics}). We found a linear evolution of the MSD with the lag time, independently of the probed location in the microfibril, supporting a standard water diffusion regime (see figure S3). The self-diffusion coefficients are computed using \reviewfirst{the slope of the linear part of} the evolution of the MSD (denoted $D$) or directional (denoted $D_{xx}$, $D_{yy}$ and $D_{zz}$) MSD from Einstein's relation. We observe an overall diffusion of water smaller than that of bulk water. Although, bulk water diffusion values are recovered at high hydration levels ($<200$\%) for $D$. In the $z$ direction, we even encounter higher self-diffusion $D_{zz}$ than in bulk water (above $2.29 \times 10^{-9}$ m$^2$.s$^{-1}$), peaking at almost $3.0 \times 10^{-9}$ m$^2$.s$^{-1}$ for high hydration levels ($> 175$\%). \reviewsecond{Such an apparent high self-diffusion coefficient could be caused by internal pressure gradients}. We witness higher water self-diffusion in the gap region in comparison to the overlap region, systematically, independently of the probed direction, at low hydration levels. The difference in self-diffusion coefficient can reach $200$\% at lowest hydration levels, varying from $0.5\times 10^{-9}$ to $1.5\times 10^{-9}$ m$^2$.s$^{-1}$. The difference in self-diffusion coefficient between the overlap and gap regions is inversely correlated with the hydration level \reviewfirst{see inset, Figure \ref{fig:water_dynamics}, the difference seems to stabilise around $2.5 \times 10^{-10}$ m$^2$.s$^{-1}$ above $200$\% hydration}. \reviewfirst{At high hydration, the sharp change in self-diffusion coefficient around $300$ \AA, at the transition between the overlap and the gap region is smoothed.}\\

\begin{figure}
    \centering
    \includegraphics[width=\textwidth]{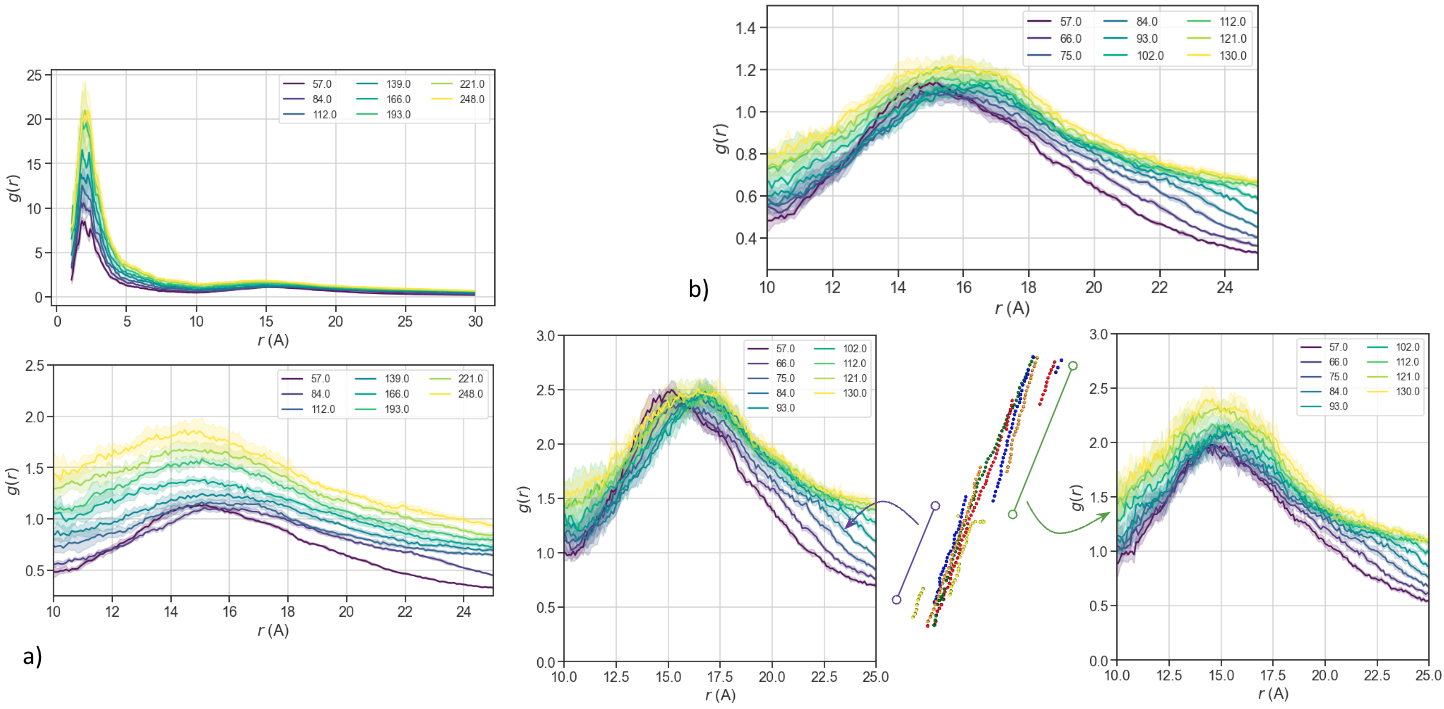}
    \caption{\reviewfirst{\textbf{Protein structure}. Evolution with hydration of the radial distribution function (RDF or $g(r)$) of the coarse-grained model of the tropocollagen molecules using one bead per triplet of amino acid. (a) Whole hydration range and (b) focus on hydration ranging from $57$\% to $130$\%, which the two bottom graphs specify the RDF in the overlap (left) and gap (right) regions. $g(r)$ is computed from a coarse-grained model of the tropocollagen helices, whereby one particle is used per triplet of amino acids (positioned at the mass centre of the amino acids of the three chains).} \reviewfirst{Error bars are obtained from time-averaging over the last $2$ ns of the simulation and from ensemble-averaging over $6$ replicas.}}
    \label{fig:protein_structure}
\end{figure}

\reviewfirst{Meanwhile, we investigate the evolution of the packing of the tropocollagen molecules, and in particular intermolecular distances (see Figure \ref{fig:protein_structure}). To that extent, we create a coarse-grained representation of the tropocollagen molecules. A particle is assigned to triplets of amino acids positioned at the same longitudinal coordinate along the tropocollagen molecule, one on each helix constituting the tropocollagen molecule. We compute the RDF of these coarse-grained particles. This coarse-graining step is introduced to reduce the information contained in the RDF spectrum. We observe a first peak at $3$ \AA~which correspond to the average distance between amino acids triplets (or between coarse-grained particles) along a tropocollagen molecule (see Figure \ref{fig:protein_structure}.a, top). We observe a second peak around which corresponds to the average distance between the centres of the tropocollagen triple helices within the fibril (see Figure \ref{fig:protein_structure}.a, bottom). This second peak is located at $14$ \AA~at $60$\% hydration. With increasing hydration, the second peak becomes rounder. Such change in the shape of the second peak indicates more variability in the distance between tropocollagen molecules. In order to better understand structural changes in the fibril assembly with hydration, we provide the RDF at smaller hydration change intervals between $60$\% and $130$\% hydration (see Figure \ref{fig:protein_structure}.b, top). We observe that the second peak, before getting rounder, shifts first from $14$ \AA~to $16.5$ \AA. The second peak can also be considered to dissociate in two distinct peaks at $14$ \AA~and $16.5$ \AA. We also isolate these RDF specifically for the overlap and gap regions (see Figure \ref{fig:protein_structure}.b, bottom). In the overlap region, the second peak located is also observed to shift to larger intermolecular distances, roughly from $15$ \AA~to $16.5$ \AA, but does not significantly get rounder. In the gap region, on the contrary, the second peak is not seen to shift but gets wider while increasing hydration from $60$\% to $130$\%. In turn, in the process of hydrating the microfibril up to $130$\%, the overlap region maintains its original packing order, only increasing intermolecular spacings, while the gap region already tends to become amorphous.}

\begin{figure}
    \centering
    \includegraphics[width=\textwidth]{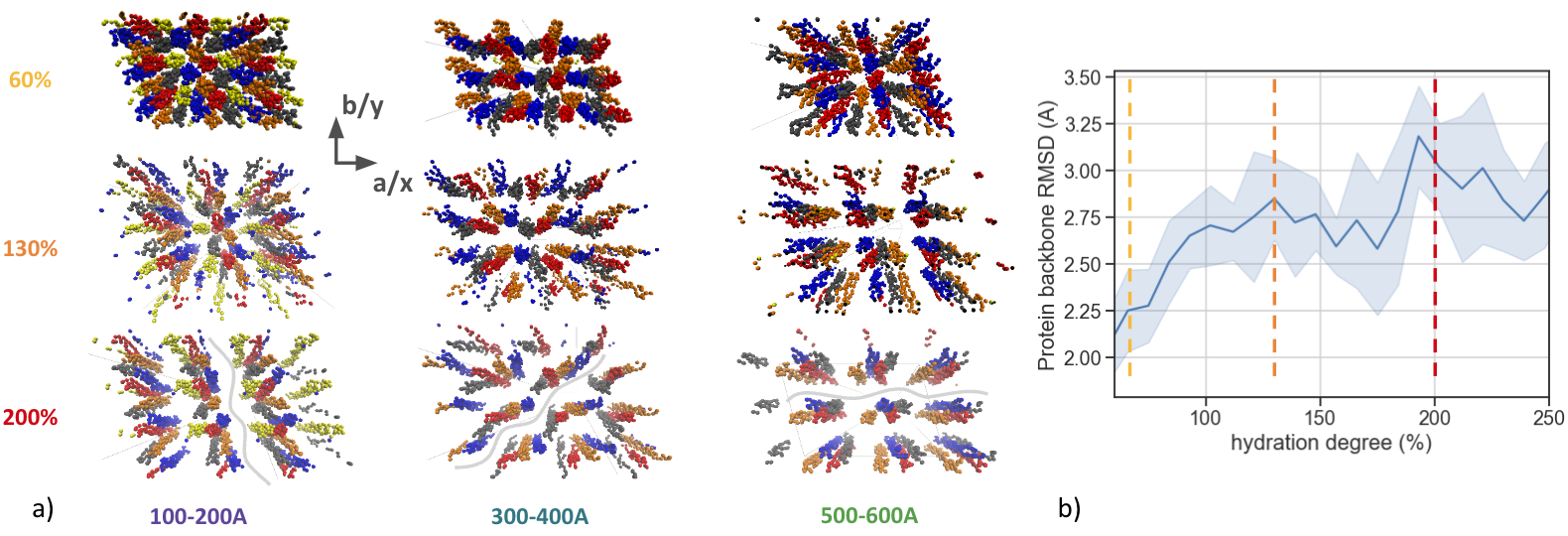}
    \caption{\reviewfirst{\textbf{Destabilization of the protein}. (a) Visualisation of the tropocollagen molecules in the microfibril at three hydration levels ($60$\%, $130$\% and $200$\%) in different cross-sections along the z-axis (z=$[100,200]$ \AA, z=$[300,400]$ \AA~and z=$[500,600]$ \AA). Hydration creates diffusion planes rotating around the long (z/c)-axis of the microfibril crystal. Empty volumes are associated with the space occupied by water molecules. Each cross section contains 3x3 crystal unit cells are shown. Continuous free volumes are highlighted in grey to guide the eye. (b) Evolution of the tropocollagen root mean squared displacement (RMSD) with the hydration degree.} \reviewfirst{Error bars are obtained from time-averaging over the last $2$ ns of the simulation and from ensemble-averaging over $6$ replicas.}}
    \label{fig:structural_visualisation}
\end{figure}

Complementary to this quantitative investigation of RDFs, we inspect qualitatively the evolution of the structure of the tropocollagen molecules within cross sections of the microfibril crystal for various $z$-axis positions at low ($60$\%), medium ($130$\%) and high ($200$\%) hydration levels (see Figure \ref{fig:structural_visualisation}.a). As we removed water molecules from the visualisation, voids correspond to the locations where water molecules localise. \reviewfirst{At low hydration, voids are only visible in the z=$[300,400]$ \AA~and z=$[500,600]$ \AA cross-sections. In turn, water molecules localise in the gap area (i.e. between $300$ and $670$ \AA~along the z-axis), in small and non percolating (discontinuous) volumes. At medium hydration, voids begin to appear also in the z=$[100,200]$ \AA cross-section, at the core of the overlap region. In contrast, at high hydration, large voids are observed, forming planes crossing continously the microfibril crystal. The planes formed by layers of water molecules have different orientations in the overlap and gap region. In the overlap region (z=$[100,200]$ \AA cross-section), the normal direction to the plane is along the $x$-axis, while in the gap region (z=$[100,200]$ \AA cross-section), the normal direction is along the $y$-axis.} \reviewfirst{We also compute the root mean squared displacement (RMSD) of $\alpha$-carbon atoms in the tropocollagen backbone (see Figure \ref{fig:structural_visualisation}.b). The RMSD of the tropocollagen molecules increases with hydration degree from $2$ to $3$ \AA. Two abrupt increases are observed $80-90$\% hydration and around $190$\% hydration.}

\begin{figure}
    \centering
    \includegraphics[width=\textwidth]{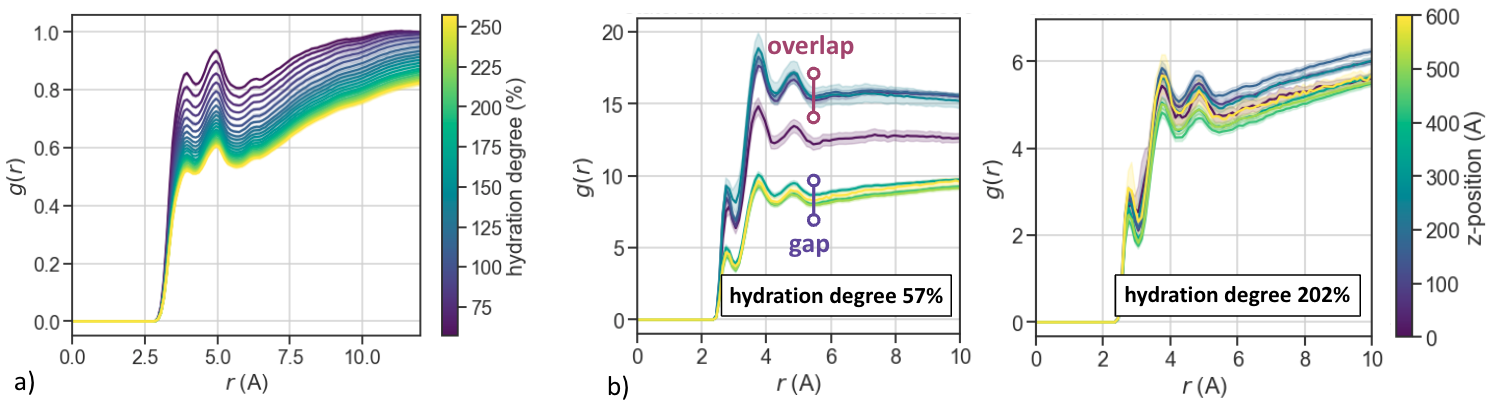}
    \caption{\textbf{Water structure around the collagen microfibril.} (a) Evolution of the RDF of water molecules around tropocollagen molecules with hydration for the whole microfibril. (b) Evolution of the RDF at (left) $57$\% and (right) $202$\% hydration degree with the position along the z-axis (slices of $3$ nm). \reviewfirst{Error bars are obtained from time-averaging over the last $2$ ns of the simulation and from ensemble-averaging over $6$ replicas.}}
    \label{fig:water-protein_structure}
\end{figure}

In order to characterise the interface between water and tropocollagen molecules, we investigate the RDF of water molecules around tropocollagen molecules (see Figure \ref{fig:water-protein_structure}). \reviewfirst{The RDF is computed between $\alpha$-carbons in the protein backbone and oxygen atoms in water molecules.} Overall in the microfibril, water molecules are less densely packed than in bulk water. Further, we observe three distinct peaks which magnitude gradually decays away as hydration increases (see Figure \ref{fig:water-protein_structure}.a), in particular the third peak. We then characterise the evolution of the structure of water molecules around tropocollagen molecules at different positions along the $z$-axis (see Figure \ref{fig:water-protein_structure}.b). Note that we compute the RDF between $\alpha$-carbons and oxygen atoms in slices of $30$ \AA~of the microfibril which causes the RDF to not converge at $1$ for large $r$ separation values. At low hydration, the three peaks are clearly observed in the core of the overlap region ($50$ to $250$ \AA) while they appear already smaller in the gap region ($350$ to $550$ \AA). At high hydration, the amplitude of the three peaks is smaller in both regions and collapses to similar values, implying that the loss of structure is much more significant in the overlap region.\\

\begin{figure}
    \centering
    \includegraphics[width=\textwidth]{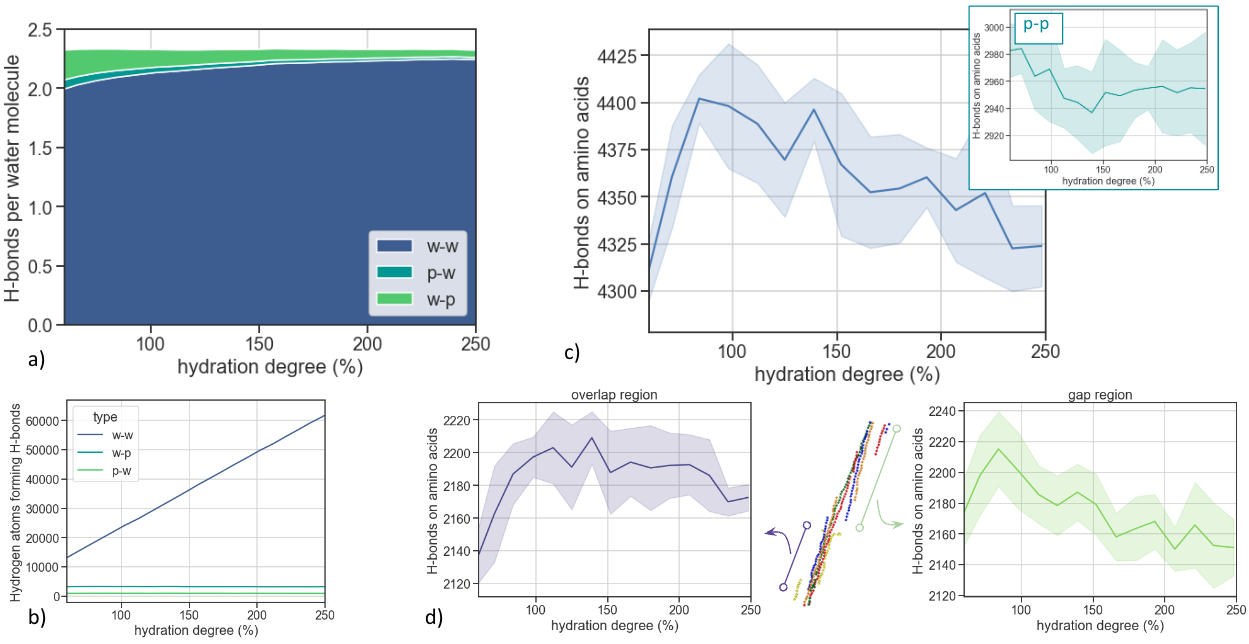}
    \caption{\textbf{Hydrogen bonds within the hydrated collagen microfibril.} Evolution with the hydration level of (a) the number of hydrogen bond per water molecule and of (b) unique hydrogen bonds (hydrogen atoms forming an hydrogen bond). A distinction is made between hydrogen bonds for which donor and acceptor are water molecules (w-w), acceptor is an atom from an amino acid (w-p) and donor is an atom from an amino acid (p-w). (c) Evolution of the number of hydrogen atoms forming hydrogen bonds with atoms from amino acids with the hydration level, the inset shows the base ground counting only hydrogen bonds with donor and acceptor being amino acids atoms (p-p); (d) detailed analysis for the gap and overlap regions. \reviewfirst{Error bars are obtained from time-averaging over the last $2$ ns of the simulation and from ensemble-averaging over $6$ replicas.}}
    \label{fig:hydrogen_bonds}
\end{figure}

We now turn to the analysis of hydrogen bonding to obtain information about the stability of the hydrated collagen microfibril (see Figure \ref{fig:hydrogen_bonds}). The analysis of hydrogen bonds involving at least one water molecule is reported in Figure \ref{fig:hydrogen_bonds}.a,b. Overall, the count of hydrogen bonds between two water molecules (\textit{w-w}) dominates and increases linearly with the hydration level, that is with the number of water molecules added in the microfibril crystal (see Figure\ref{fig:hydrogen_bonds}.b). In comparison, the number of hydrogen bonds between water molecules and the amino acids of the collagen protein remains stable. Further, we observe a stable number of hydrogen bonds per water molecules in the system (see Figure \ref{fig:hydrogen_bonds}.a) around $2.3$. This is less than the number of hydrogen bonds per water molecule in bulk water which is about $3.3$ \cite{smith_energetics_2004}. We then focus on hydrogen bonds formed within and at the surface of the tropocollagen molecules (see Figure \ref{fig:hydrogen_bonds}.c,d). We observe that the number of hydrogen bonds in between protein amino acids  (\textit{p-p}) fluctuates but remains rather constant throughout the hydration process (see inset in Figure \ref{fig:hydrogen_bonds}.c). These protein-protein hydrogen bonds are formed almost exclusively within $\alpha$-helix chains and between triplets of helices forming tropocollagen molecules. And rather rarely between distinct tropocollagen molecules. The stability of the protein-protein hydrogen bonds count demonstrates that the individual tropocollagen structure is not altered by the hydration level. We confirm this showing that the survival of protein-protein hydrogen bonds is much longer than hydrogen bonds involving at least one water molecule (see Figure S5). Now, observing the hydrogen bonds that involve at least one amino acid atom, the evolution with the hydration level is not as much stable (see Figure \ref{fig:hydrogen_bonds}.c), a peak or a plateau can be seen around $90$\% hydration. We looked whether this variation in hydrogen bond count held in the gap and overlap regions of the crystal (see Figure \ref{fig:hydrogen_bonds}.d). The bond count appeared rather stable in the overlap region up to $200$\% hydration, while a decay is observed in the gap region at much lower hydration.\\

\section{Discussions}
\label{sec:discussions}

\subsection{Water structure and dynamics}

Hydration dictates the structure and dynamics of water molecules. This structure becomes more ordered as hydration increases as shown by the narrower peaks in radial distribution functions. The broader second peak, hardly discernible from the first peak, at low hydration implies that the structure of water is far from bulk structure, whereby water displays medium-range order associated with three peaks \cite{wang_density_2011} and tetrahedral organisation \cite{rahman_molecular_2003, renou_water_2014}. Medium-range order is only recovered at high hydration level (beyond $150$\% hydration) indicating localisation of bulk water between protein molecules. This water-water organisation may imply that water molecules are initially found in layers confined between tropocollagen molecules, but only at low hydration. Looking at the organisation of water molecules around tropocollagen molecules, we do also observe three peaks which the two first correspond to experimentally observed peaks at $2.7$~\AA~and $3.6$~\AA~\cite{berisio_crystal_2002}. This organisation led to the water mono-layer model surrounding tropocollagen molecules proposed by Fullerton et al.\cite{fullerton_evidence_2006} Nonetheless, the amplitude of these peaks decays with the hydration level, supporting the recovery of bulk water structure and dynamics at high hydration.

The diffusion of water in the microfibril crystal is overall normal (see figure S3), and in particular in the $z$ direction. Yet, subdiffusive regimes may be expected as they have been reported at the interface with proteins using neutron scattering experiments \cite{doster_proteinwater_2005}. Subdiffusive regimes correspond to $MSD \propto t^{\alpha}$ with $\alpha < 1$ \cite{gallo_anomalous_2003}. Looking in more detail at the diffusion exponents in different locations of the microfibril crystal along the $z$-axis and at varying hydration levels (see respectively figures S3.a and S3.b), we found that in the overlap region, in particular close to the dense region where telopeptides localise, and at low hydration level, the diffusion exponent drops. The exponent decreases below $0.9$ when focusing on the MSD along the $x$ or the $y$ axes. Therefore, in conditions associated with high confinement we do observe anomalous diffusion. This is consistent with former computational results of water dynamics implying that anomalous diffusion at the interface with proteins decreases when protein molecules regain mobility \cite{pizzitutti_how_2007}. Further, similarly to the recovery of bulk water medium-range order, the self-diffusion coefficient of bulk water is recovered in the microfibril at high hydration level. Last, in the dense overlap region, at low hydration level, water self-diffusion is as low as for measurements in supercooled water below $260$ K \cite{pruppacher_selfdiffusion_1972}.

\subsection{Collagen structure and dynamics}

\reviewfirst{The insertion of water molecules in the microfibril crystal modulates the tropocollagen intermolecular spacings and forces. At physiological hydration, about $60$\%, we simulated an average intermolecular spacing of $1.4$ nm. Such value is close to but slightly overestimating experimental observations about $1.3$ nm \cite{lees_water_1986, morin_fibrillar_2013}. One explanation may lie in cross-linking, which is not considered in the present study. Cross-linking may oppose the swelling of the fibril, especially in mature tissues showing high degree of reticulation. This may imply an overestimation of intermolecular spacings. Nonetheless, the first origin of cross-linking is enzymatic, binding adjacent telopeptides of tropocollagen molecules. In the present microfibril simulations, this would only lead to the addition of one cross-link in the middle of the overlap region, between two out of five tropocollagen strands. In turn, we do not expect that the addition of enzymatic cross-link would strongly affect the simulated protein structure, but could potentially affect intermolecular spacings.}

\reviewfirst{Further, our simulation predict a $1.5$-fold volume increase following a $50$\% hydration increase (see Figure S1). This is consistent with experimental data following a similar increase in hydration, although at different absolute hydration levels \cite{grant_tuning_2009, andriotis_hydration_2019}. Yet, our simulations overestimate other experimental observations which found a $1.2$-fold volume increase \cite{spitzner_nanoscale_2015}. Comparison with experimental data on swelling is made difficult by the constraining effect of the substrate in experimental measurements and different hydration conditions usually moving from air to buffer, that is from $30$\% to $80$\% hydration levels \cite{spitzner_nanoscale_2015}.}

\reviewfirst{With increasing hydration, the observed multiple intermolecular distances and their increased variability may correspond to anisotropic swelling of the microfibril crystal. In particular, an anisotropic dilatation of the crystal along the $x$-axis and $y$-axis. From the snapshots of the structure at high hydration, in the overlap region, it seems that the spacings along the $x$-axis are larger than along the $y$-axis; and in the gap region, it seems to be the inverse. In both cases, hydration causes the protein structure to display anisotropic intermolecular spacings between $100$\% and $130$\% hydration, before becoming fully amorphous beyond $130$\%. These visual inspections of the microfibril are confirmed by the site-specific RDF in the overlap and gap regions. The overlap region displays increased intermolecular spacings and a preserved packing order, while the gap region already becomes amorphous, losing quasi-hexagonal packing \cite{zhou_supramolecular_2016}, showing a higher variability of intermolecular spacings. This is consistent with the heterogeneous swelling of collagen fibrils reported by Spitzner et al. \cite{spitzner_nanoscale_2015}, that is a difference in swelling between the gap and overlap region.}

\reviewfirst{Swelling appears to be not only variable along the $z$-axis but also anisotropic, occurring along a plane (grey lines in Figure \ref{fig:structural_visualisation}.a). This is particularly visible at $200$\% hydration. The normal direction to the plane is along the $x$-axis in the overlap region and along the $y$-axis in the gap region. This seem to indicate that the plane is rotating around the $z$-axis as we move from the overlap to the gap region. These quantitative observation tend to support the earlier observation that the microfibrils swells with hydration, displaying water two-dimensional water transport channels.}

\subsection{Stability and disassembly of hydrated fibril}

\reviewfirst{We simulated the hydration of the collagen microfibril up to levels which could potentially induce its disassembly. The destabilisation of the microfibril seems to initiate in the gap region as early as $100$\% hydration. This can be inferred from the decrease in hydrogen bond counts but also from the increase in intermolecular spacing. Interestingly, the water structure around tropocollagen molecules is much less organised even at low hydration levels. The destabilisation of the microfibril appears then to propagate to whole fibril when reaching higher hydration levels. In the overlap region the destabilisation is more progressive than in the gap region, as seen from the loss of structure of water molecules around tropocollagen molecules. Hydration causes the protein structure to become fully amorphous beyond $130$\% which could support the onset of disassembly. At such hydration, planes formed across the whole microfibril tend to appear, while the flexibility of tropocollagen molecules significantly increases. Meanwhile, the sharp gradient of the self-diffusion coefficient of water molecules between the gap and overlap region is smoothed. However, bulk water diffusion is only recovered at hydration levels beyond $200$\%.\\} 

\reviewfirst{Destabilization of the fibril is supported by the evaluation of the energy stored in the tropocollagen molecules (see Figure S4). Intermolecular forces correspond to electrostatic and van der Waals components decrease with hydration, except between $60$\% and $90$\% hydration, whereby an increase is observed. This observation is consistent with the evolution of hydrogen bond counts, but also with experimental observations showing that intermolecular forces tend to decrease with hydration \cite{leikin_temperature-favoured_1995}. This also supports an earlier finding on the role of water molecules in collagen fibril, which states that water molecules act at times (at low and medium hydration levels) as a "glue" and at times (at high hydration levels) as a "lubricant" between tropocollagen molecules \cite{zhang_effect_2007}. Our results hint a transition between the two regimes beyond $100$\% hydration. Hydration level between $90$\% and $100$\% of water to collagen mass ratio maximises hydrogen bonding on tropocollagen molecules. This hydration range should therefore be optimal for force transfer between tropocollagen molecules mediated by water bridges in our simulations. Interestingly, this hydration level is higher than the physiological hydration level (about $62$\%) of collagen fibers in rat tail tendon \cite{fullerton_orientation_1985}, which may hint that collagen fibrils are not solely designed to \reviewsecond{draw optimal mechanical potential from hydrogen bonds.}}

\subsection{Mechanical properties and biomineralisation}

Extrapolating to the hydrated microfibril mechanics, elastic properties are expected to vary specifically with hydration. More precisely, high hydrogen bonding on the tropocollagen molecules at low to moderate hydration ($80$\% to $100$\%) will reduce sliding between the triple helices and induce highest bending modulus. Meanwhile, reduced water diffusion up to $150$\% hydration will impede water escape during longitudinal ($z$-axis) stretching of the microfibril, rendering it incompressible, and therefore maximise it Young modulus.\\

Water dynamics may also enlighten us about the biominerlisation of the microfibril. The water dynamics are highly anisotropic in the collagen crystal. Even more so, in the gap region. While diffusion along the $z$-axis largely dominates at low and moderate hydration levels, the rat tail tendon physiological hydration, diffusion also varies between orthogonal $x$ and $y$ axes. Diffusion along the $x$-axis is higher than along the $y$-axis. We hypothesise that this anisotropy of the diffusion coefficient could induce an heterogeneous mineral ion diffusion within the microfibril, and therefore be responsible for the anisotropic shape of the hydroxyapatite crystals found in mineralised collagen, often reported as two-dimension platelets \cite{xu_intermolecular_2020}. Subsequently, the assumption that higher diffusion leads to higher transport of mineral ions, the mineral platelets should preferably aligned with the $z$-axis of the microfibril crystal. And the second largest dimension of the platelets should be along the $x$-axis, which is aligned with the radius of collagen fibrils \cite{perumal_collagen_2008}. Consequently, the hydroxyapatite platelets should be oriented radially within mineralised collagen fibrils, favoring observations of interwoven organic and mineral phases \cite{reznikov_fractal-like_2018}, but disagreeing with hypothesis of randomly oriented platelets in the collagen fibril cross-section \cite{xu_intermolecular_2020}. 

\section{Conclusions}
\label{sec:conclusion}

We built a molecular model of the hydrated collagen microfibril. \reviewfirst{We performed simulations of the hydration of microfibril up to $250$\% in order to trigger the onset of disassembly.} We investigated the variations of water structure, water dynamics and hydrogen bonding with hydration and with spatial resolution. We found that water is tightly confined at low hydration levels, whereby water molecules are organised in mono-layers and exhibit slow diffusion, while at high hydration levels water molecules recover bulk water properties. \reviewfirst{We also characterised the structure and dynamics of tropocollagen molecules within the fibril. We found that the destabilization of the protein structure occurs at lower hydration degrees in the gap region than in the overlap region, which rapidly becomes amorphous. The protein structure displays increasing anisotropy with increasing hydration, with water molecules localising in a continuous plane rotating around the long axis of the crystal.} Overall, the role of water confined in the collagen microfibril changes with hydration, reinforcing interactions between tropocollagen molecules below $100$\% hydration while favouring sliding and disassembly beyond $130$\% hydration. Our findings highlight that sharp control over hydration of collagen fibrils enables organisms to switch from assembly to disassembly and to tune the mechanical properties of the assembled tissue. In turn, our molecular dynamics simulations provide new insights to understand the contribution of water molecules to the structure-property relationships of collagen fibrils.

\section{Acknowledgements}
\label{sec:acknowledgements}

The author acknowledges funding support from the Fondation ARC through its postdoctoral fellowship program. The author gratefully acknowledges the UCL Advanced Research Computing centre (www.ucl.ac.uk/advanced-research-computing) for providing computing time on the Kathleen supercomputer. The author thanks Shunzhou Wan for helping with setting up molecular dynamics simulations with NAMD and Denis Morineau for the fruitful discussions on water structure and diffusion and on hydrogen bonding.

\section{Supporting Information}

The supporting information contains supplementary figures: (i) the evolution of the water diffusion exponent to characterise the diffusion regime in the collagen microfibril; (ii) the radial distribution of protein backbone atoms as well as the lifetime of protein hydrogen bonds to characterise the structure of the protein phase of the microfibril; (iii) the variation in water structure around the tropocollagen backbones to characterise the water-protein interface. The data serves as a validation of the molecular model.

\section{Data availability}

The structures of the hydrated microfibril will be made available on the \textit{Recherche Data Gouv Repository} upon publication.

\section{Code availability}

The scripts that were used to build the molecular model of the microfibril and to simulate the iterative hydration process will be made available on the \textit{Recherche Data Gouv Repository} upon publication.


\section{Ethics declaration}

The Author declare no Competing Financial or Non-Financial Interests.

\newpage

\bibliography{references.bib}

\newpage

\begin{figure}
    \centering
    \includegraphics[width=\textwidth]{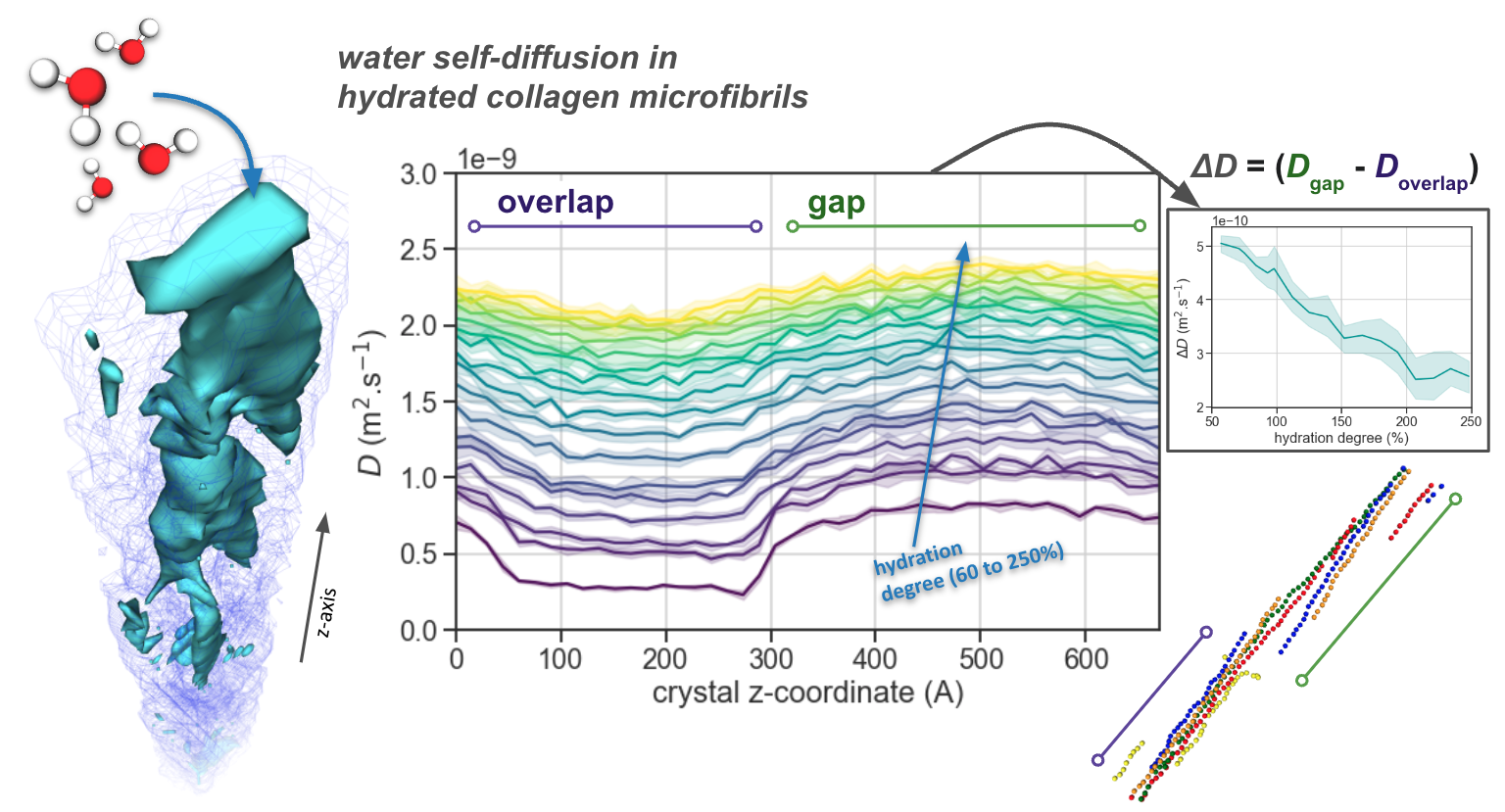}
    \caption{\textbf{For Table of Contents Use Only.}}
    \label{fig:toc}
\end{figure}

\end{document}


\maketitle



\break

\begin{figure}
    \centering
    \includegraphics[width=\textwidth]{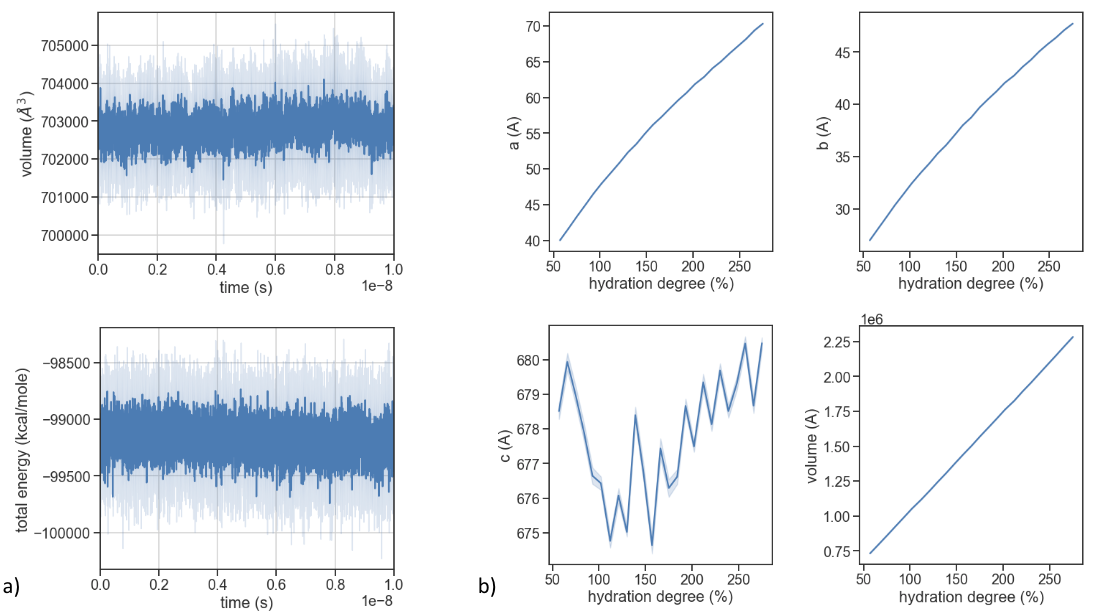}
    \caption{\textbf{The hydrated collagen microfibril crystal.}  (a) Evolution of the the energy and volume of the hydrated microfibril after initial hydration with $12550$ water molecules. (b) Evolution of the dimensions and volume of the microfibril crystal at different hydration levels.}
    \label{fig:equilibration_hydration}
\end{figure}

\break

\begin{figure}
    \centering
    \includegraphics[width=\textwidth]{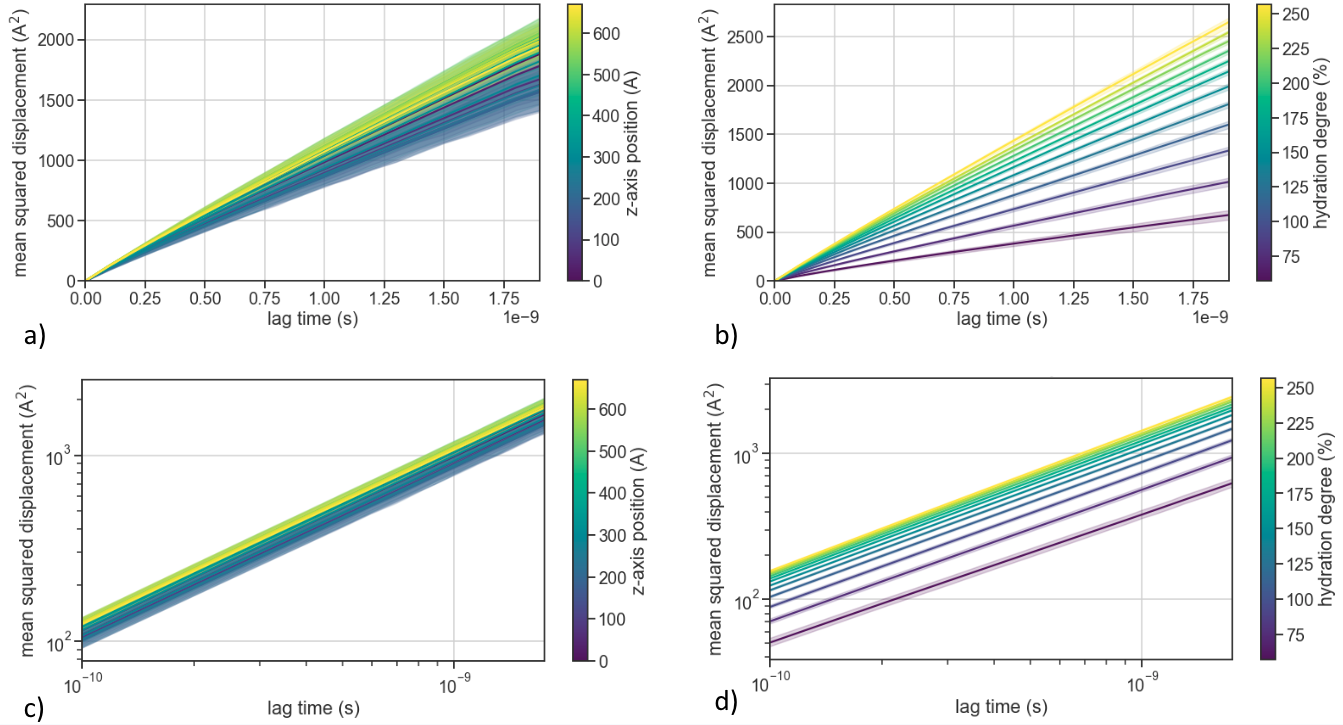}
    \caption{\textbf{Mean squared displacement of water molecules.} (a,c) Evolution of the MSD with the lag time as a function of the long z-axis position in the microfibril crystal. (b,d) Evolution of the MSD with the lag time as a function of the hydration degree. (a,b) Plot in linear-linear scale. (c,d) Plot in log-log scale.}
    \label{fig:water_msd}
\end{figure}

\break

\begin{figure}
    \centering
    \includegraphics[width=\textwidth]{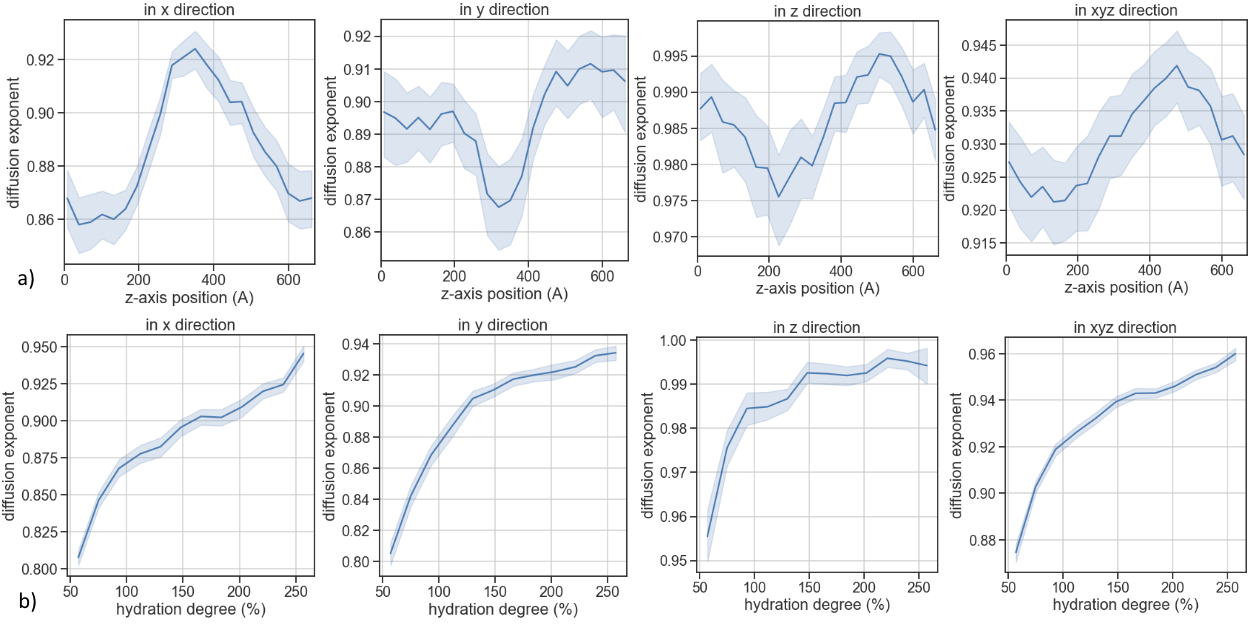}
    \caption{\textbf{Looking for anomalous diffusion.} Looking for anomalous diffusion. Evolutions of the diffusion exponent (slope of the log-log MSD vs lagtime plots) with (a) the position in the microfibril crystal along the long z-axis and (b) the hydration level. The diffusion exponent is computed for each MSD direction (x, y or z) and for the magnitude of the MSD (xyz).}
    \label{fig:diffusion_exponent}
\end{figure}

\break

\begin{figure}
    \centering
    \includegraphics[width=\textwidth]{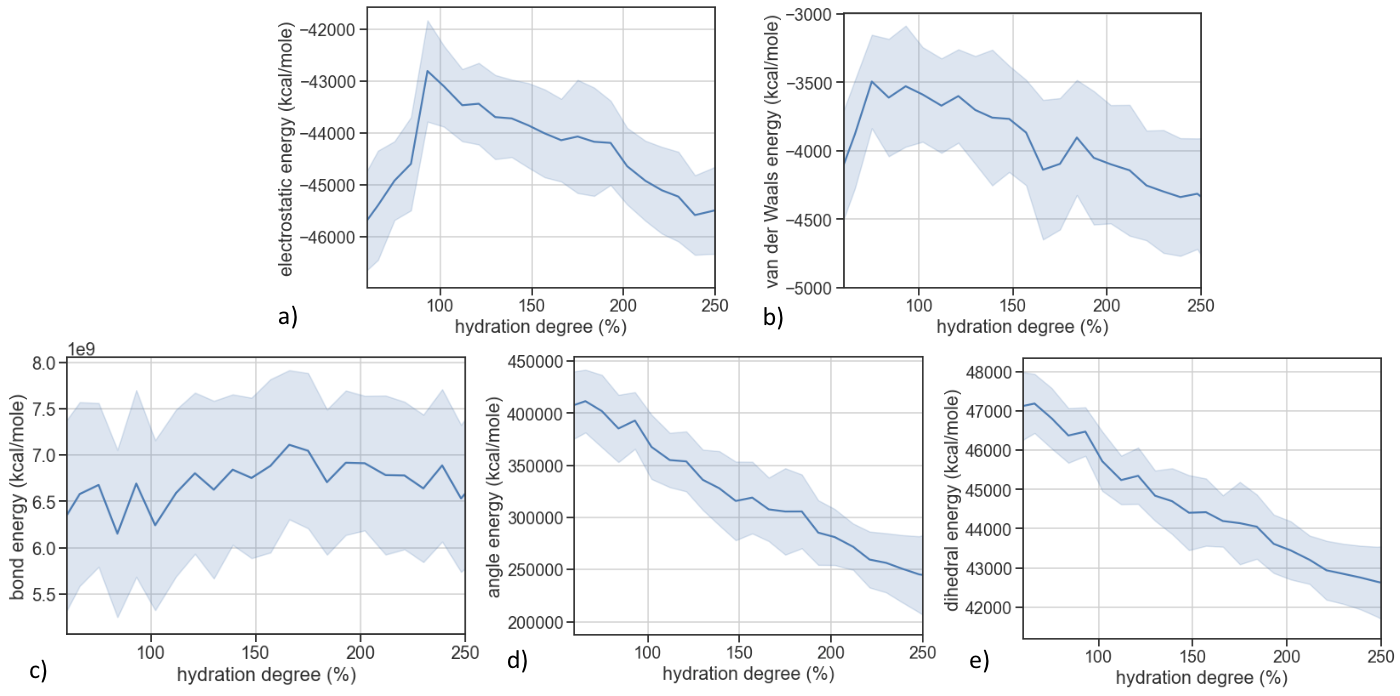}
    \caption{\textbf{Protein potential energy.} Evolution of the elementary contributions of the potential energy of the tropocollagen molecule with the hydration degree: (a) electrostatics, (b) van der Waals, (c) covalent bonds, (d) three-body angles, and (e) dihedral angles. Error bars correspond to time and ensemble-averaging}
    \label{fig:protein_energy}
\end{figure}

\break

\begin{figure}
    \centering
    \includegraphics[width=0.8\textwidth]{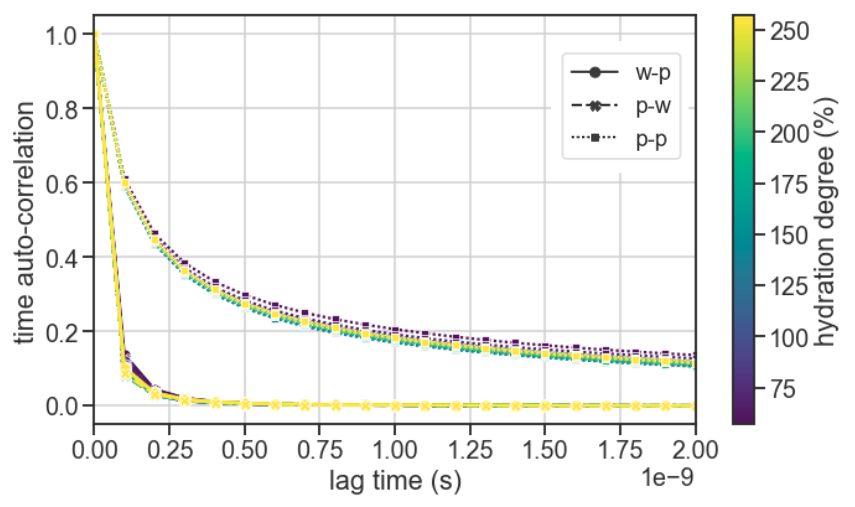}
    \caption{\textbf{Hydrogen bond life-time.} Stability (time auto-correlation) of hydrogen bonds formed on the protein atoms with other protein atoms ('p-p') or with water atoms ('w-p' or 'p-w') for varying degree of hydration; we see no influence of the hydration, but protein-protein hydrogen bonds are more stable than water-protein ones.}
    \label{fig:protein_characterisation}
\end{figure}
